\documentclass[letter]{aa}

\usepackage{natbib}
\usepackage{graphicx}
\usepackage{newtxtext,newtxmath}
\usepackage{txfonts}
\usepackage{float}
\usepackage{mathrsfs}
\usepackage{amsmath}
\usepackage{mathtools}

\newcommand{\dg}{$^{\circ}$}
\newcommand{\rsun}{$R_{\odot}$}

\newcommand{\tiii}{type~{{{III}}}}
\newcommand{\tj}{type~{{{J}}}}
\newcommand{\tu}{type~{{{U}}}}
\newcommand{\Tu}{Type~{{{U}}}}
\usepackage{color}

\begin{document}

\title{Three-dimensional  reconstruction of \tu\ radio bursts: \\ a novel remote sensing approach for coronal loops}
\author{S. Mancuso\inst{1}, D. Barghini\inst{1,2}, A. Bemporad\inst{1}, D. Telloni\inst{1}, D. Gardiol\inst{1}, F. Frassati\inst{1}, I. Bizzarri\inst{2} \and C. Taricco\inst{1,2}}

\institute{
${^1}$ Istituto Nazionale di Astrofisica, Osservatorio Astrofisico di Torino, via Osservatorio 20, Pino Torinese 10025, Italy \\ 
${^2}$  Università degli Studi di Torino~--~Dipartimento di Fisica, via Pietro Giuria 1, Torino (TO), Italy \\ \email{salvatore.mancuso@inaf.it}}
\date{Received \today\ / Accepted}

\abstract{
\Tu\ radio bursts are impulsive coherent radio emissions produced by the Sun that indicate the presence of subrelativistic electron beams propagating along magnetic loops in the solar corona. 
In this work, we present the analysis of a \tu\ radio burst that was exceptionally imaged on 2011 March 22 by the Nançay Radioheliograph (NRH) at three different frequencies (298.7, 327.0, and 360.8 MHz).
Using a novel modelling approach, we show for the first time that the use of high-resolution radio heliograph images of \tu\ radio bursts can be sufficient to
both accurately reconstruct the 3D morphology of coronal loops (without recurring to triangulation techniques) and to fully constrain their physical parameters.
At the same time, we can obtain unique information on the dynamics of the accelerated electron beams, which provides important clues as to the plasma mechanisms involved in their acceleration and as to why \tu\ radio bursts are not observed as frequently as \tiii\ radio bursts.
We finally present plausible explanations for a problematic aspect related to the apparent lack of association between the modeled loop as inferred from radio images and the extreme-ultraviolet (EUV) structures observed from space in the same coronal region.

}

\titlerunning{3D reconstruction of \tu\ radio bursts}
\authorrunning{Mancuso et al.}
\maketitle

\section{Introduction}

Solar \tu\ radio bursts, first discovered by \cite{Maxwell1958}, represent rare spectral features characterized by an inverted-U shape on radio dynamic spectra.
The emission mechanism is supposed to be similar to that responsible for the emission of the more commonly observed \tiii\ radio bursts previously reported by \cite{Wild1950}.
The latter is the most common coherent  solar radiation produced via plasma emission at the fundamental and/or second harmonic by subrelativistic energetic beams of electrons that travel along open magnetic field lines at distances up to 1 AU.
The basic mechanism for \tiii\ radio emission, originally proposed by \cite{Ginzburg1958}, involves wave--wave processes with ion-sound waves for the fundamental emission, and with oppositely directed Langmuir waves for second harmonic emission (see the recent review from \citealt{Reid2020}).
As these high-energy electron beams propagate through the coronal plasma with decreasing background electron densities, and hence decreasing plasma frequency, they emit \tiii\ radio emission at correspondingly decreasing radio frequencies. 
Their analysis can therefore provide useful insight into the astrophysical processes pertaining  to the acceleration and transport of energetic charged particles of solar origin and also to the physical properties of the plasma through which they propagate.
On the other hand, not all electron beams that generate \tiii\ radio bursts are able to escape into the solar wind.
Some of them are compelled to travel along confined coronal magnetic loops, thus forming inverted U shapes in radio dynamic spectra.
Moreover, being constrained by the loop's magnetic field, the electron beams, after reaching the top of the loop, travel down the opposite leg through an increasing plasma density, thus producing a positively drifting frequency rate.
Inverted \tu\ bursts are therefore bounded by the turning frequencies reached at the apex of the loop that might occur at MHz (\citealt{Stone1971}) or even GHz frequencies (\citealt{Wang2001}).
As most of the solar magnetic flux in the corona has a closed-loop configuration,  it is somewhat surprising that \tu\ radio bursts are not observed as frequently as \tiii\ radio bursts. 
This could be related to the peculiar mechanism that limits the production of \tu\ radio bursts with respect to \tiii\ radio bursts.
According to \cite{Reid2017}, the rarity of type U bursts could be due to the fact that the accelerated electron beams need to travel a certain distance along the loop to become unstable to Langmuir waves, which in most cases is greater than the linear length of the loops.
In some cases, the descending branch of \tu\ radio bursts presents an abrupt interruption, resulting in \tj\ radio bursts: this could be a consequence of the energy loss of the emitting beam before covering the descending portion of the coronal loop (\citealt{Fokker1970}).
\Tu\ radio burst analyses are therefore a powerful and unique remote sensing diagnostic tool for characterizing both the dynamics of the exciting electron beams and the confined magnetized plasma they travel through. 

The present paper is structured as follows: in Section 2, we describe the observations of a \tu\ radio burst that was detected by ground-based radio dynamic spectra and imaged by a radio heliograph at three different frequencies, with spatial and temporal resolving powers able to separate both its ascending and descending branches. 
In Section 3, we present a novel 3D modeling approach to reconstructing the loop morphology on the basis of the observed radio signatures.
The results from this section are then used to characterize the physical properties of the coronal loop and to further constrain the characteristics of the accelerated electron beam that produced the observed coherent radio emission.
Finally, in Section 4, we discuss the results and draw our conclusions. 

 \begin{figure}
    \centering
    \includegraphics[width=8.6cm]{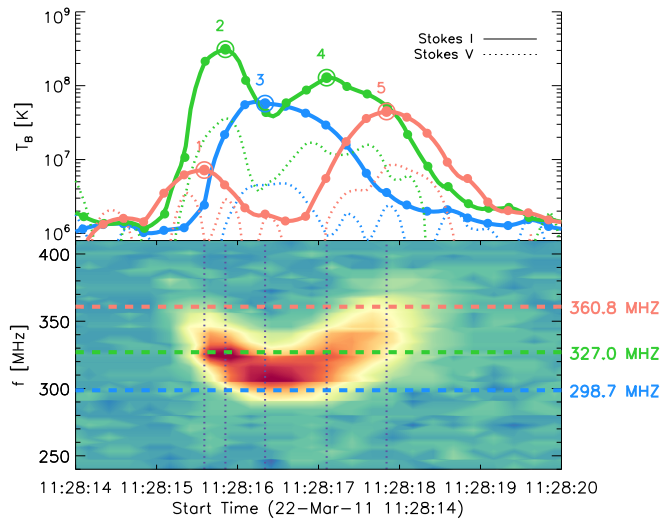}
    \caption{Spectral data collected by the e-CALLISTO-BLEN radio spectrometer during the \tu\ emission (bottom) together with temporal profiles of the radio brightness temperature $T_B$ based on NHR observations (top). The vertical lines denote the time of the NRH images corresponding to intensity maxima at each frequency.}
    \label{Fig1}
\end{figure}

\section{Observations}

The \tu\ radio burst was detected in the dynamic spectral data obtained by the Compact Astronomical Low-cost Low-frequency Instrument for Spectroscopy and Transportable Observatory (CALLISTO) BLEN radio spectrograph\footnote{http://www.e-callisto.org/.} (170--870 MHz with a cadence of 0.25 s).
The dynamic radio spectrum related to this event, taken  on 2011 March 22 over a time interval of about 6 s between 11:28:14 UT and 11:28:20 UT, is shown in the bottom panel of Figure~\ref{Fig1}.  
The \tu\ radio burst started at about 11:28:15 UT and lasted for nearly 3 seconds, occurring at frequencies between about 300 and 370 MHz. 

 \begin{figure}
    \centering
    \includegraphics[width=8.6cm]{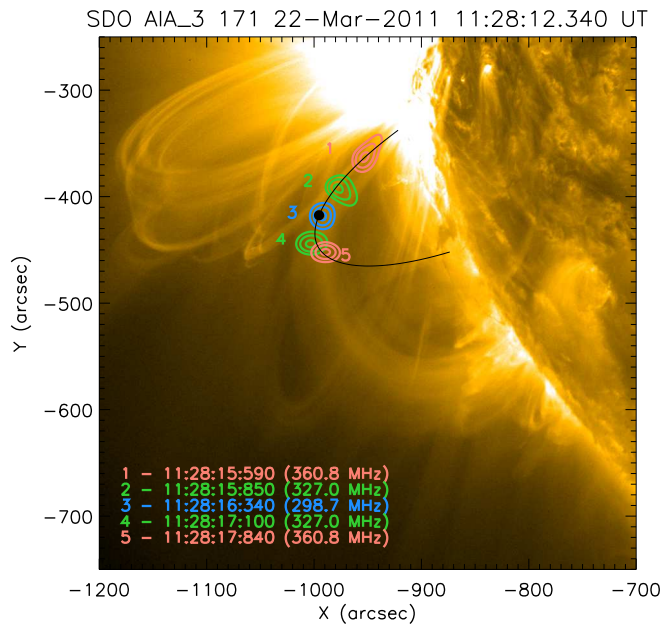}
    \caption{SDO/AIA image taken at 171 \AA\ on 2011 March 22 at 11:28 UT showing the coronal configuration during the \tu\ radio emission.
    The radio source positions were imaged by the NRH at the intensity peaks at 298.7, 327.0, and 360.8 MHz. The colored contours represent 99.0\%, 99.5\%, and 99.9\% of the corresponding intensity maximum and the three frequencies are plotted in blue, green, and red, respectively. The black solid curve represents the modeled loop, while the black filled circle indicates the apex of the loop.} 
    \label{Fig2}
\end{figure}

The propagation of the \tu\ radio emission was simultaneously imaged by the Nançay Radioheliograph (NRH; \citealt{Kerdraon1997}), apparently originating above active region NOAA 11176, which at the time of the radio emission was near the eastern limb of the Sun. 
The NRH is a synthesis instrument that provides 2D Stokes I and V images of the Sun with subsecond time resolution.
On that day, the NRH  made observations of the Sun at all ten operating frequencies (445.0, 432.0, 408.0, 360.8, 327.0, 298.7, 270.6, 228.0, 173.2, and 150.9 MHz) with a cadence of 0.25 s. 
The NRH data were obtained from the Solar Radio Data Base website\footnote{https://rsdb.obs-nancay.fr}, which provides free access to the radio observations of the Sun obtained by the NRH Radioheliograph and the two spectrographs ORFEES and NDA at the Nançay radioastronomy station in France.  
The data are distributed as binary files in a Nançay proprietary format, which contain the visibilities measured by NRH that were processed using the NRH widget of the Solarsoft system (SSW; \citealt{Freeland1998}). 
The processed data contain measurements of both the Stokes I (total intensity) and V (circular polarization) components of the radio emission, recorded as brightness temperature $T_B$ and measured in Kelvin. 
The standard technique within the SSW software was used to synthesize the time series of the 2D radio intensity maps from the NRH visibilities of $128 \times 128$ pixels with the pixel size of 30" for each NRH frequency.
At 11:28 UT, the beam size (full width at half maximum) at the three observation frequencies (298.7, 327.0, and 360.8 MHz) was 78", 71", and 64" along the major axes, and 63", 58", and 53" along the minor axes, respectively. 
The beam was almost circular due to the fact that the event time was favorable to precise observations with NRH  during the day, because of minimal zenith distance of the Sun.

\begin{figure}
    \centering
    \includegraphics[width=8.6cm]{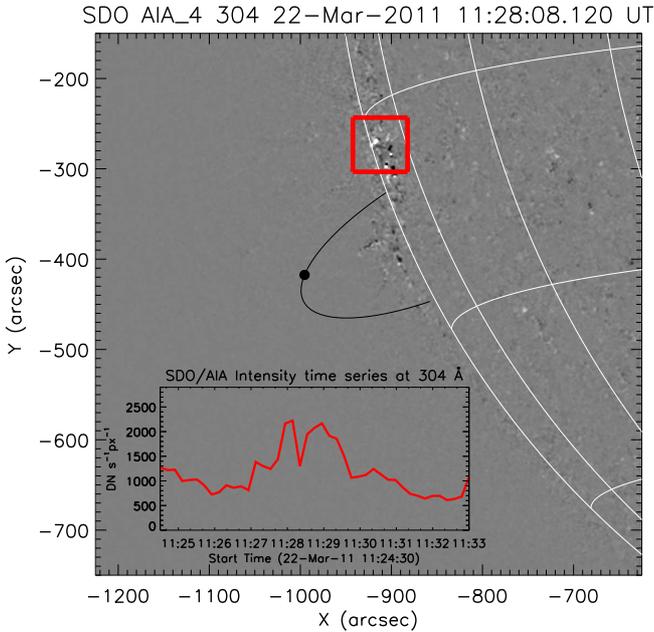}
    \caption{Running difference image from SDO/AIA at 304 \AA\ taken at 11:28 UT showing a brightening at the southeast limb, peaking around the time of the type U burst, just above the position of the modeled loop. The inset shows the temporal behavior of the maximum EUV intensity at this wavelength from about 11:24 to 11:33 UT taken in the area specified by the red square.}
    \label{Fig3}
\end{figure}

The radio burst was imaged at three frequencies (298.7, 327.0, and 360.8 MHz).
The top panel of Figure~\ref{Fig1} shows the temporal profiles of $T_{\rm B}$ maxima and circular polarization levels (i.e., Stokes I and Stokes V) observed by NRH from 11:28:14 UT to 11:28:20 UT. 
Both the ascending and descending branches of the \tu\ radio bursts have the same sign of polarization.
This can be interpreted in terms of mode coupling, which supposes the existence of the quasi-transverse region in the overlying loop systems (\citealt{Benz1977}).
The fractional circular polarization, $\rho = {\rm V/I}$, is about 10\% throughout the event.
The weak polarization observed for this event is expected because it occurred near the limb (e.g., \citealt{Suzuki1980}).
Moreover, this is consistent with previous observations of \tu\ bursts, which revealed that the degree of polarization is generally low (\citealt{Benz1977,Aschwanden1992,Aurass1997}). 
In Figure~\ref{Fig2}, we show an extreme ultraviolet (EUV) intensity map taken at 171 \AA\ by the Atmospheric Imaging Assembly (AIA; \citealt{Lemen2012}) on board the Solar Dynamics Observatory (SDO; \citealt{Pesnell2012}) superposed with the contours of the NRH radio brightness temperature ($T_{\rm B}$) of the radio burst obtained at about the same time.
Radio source positions at the intensity peaks observed at the three frequencies listed above are delineated by colored contours that represent 99\%, 99.5\%, and 99.9\% of the corresponding maximum.
As a final remark, we point out that the \tu\ radio burst investigated in this work is not associated with solar eruptive phenomena, implying that the energetic electrons emitting the coherent radio emission are most probably accelerated by nonflaring processes.
However, a weak small-scale brightening was detected in SDO/AIA images at 304 \AA\ at the southeast limb, peaking around the time of the type U burst, just above the position of the modeled loop (see Figure~\ref{Fig3}). 

 \begin{figure*}
    \centering
    \includegraphics[width=17cm]{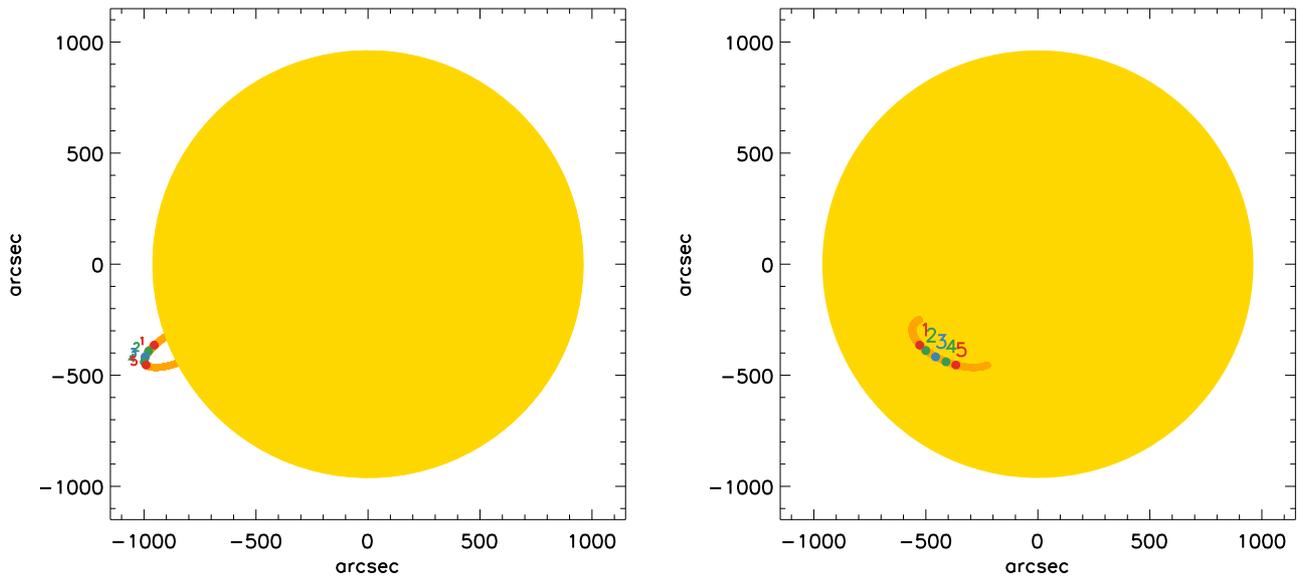}
    \caption{Schematic view of the coronal loop model as seen from the perspective of Earth (left panel) and rotated by 90\dg\ to the west (right panel). The small colored circles on the loop yield a 3D view of the  positions of the radio sources along the loop as seen by NRH (see also Figure \ref{Fig2}).}
    \label{Fig4}
\end{figure*}

\section{Data analysis and results}

\subsection{Coronal loop morphology}

In general, the reconstruction of the geometry of coronal loops is not a straightforward task since they present a variety of shapes and orientations with respect to the line of sight and are often inclined with respect to the radial direction.
On the other hand, as a good approximation, loops are expected to present a semi-elliptical shape (e.g., \citealt{Reale2014}).
In this work, we adopt a novel technique to infer the 3D morphology of the coronal loop (given the observed \tu\ signature) and the dynamics of the electron beam responsible for the coherent radio emission without utilizing triangulation techniques.
In particular, we adopt a semi-ellipse 3D loop modeling approach to match the morphology of the loop as observed by NRH in the plane of the sky.

A generic rotation in 3D of an inclined semi-elliptical loop is described by the product of successive rotations around the principal axes of the Cartesian coordinates as a rotation matrix: 
\begin{equation}
{\rm R} = {\rm R_z}{\rm R_y}{\rm R_x}{\rm R_\delta},
\end{equation}
where 
\begin{equation}
{\rm R_x}\equiv
\begin{pmatrix}
      1      &        0      &    0     \\
      0      & \cos{\gamma} & -\sin{\gamma} \\
      0      & \sin{\gamma} &  \cos{\gamma} \\
\end{pmatrix}, \hspace{0.2cm}
{\rm R_y}\equiv
\begin{pmatrix}
\cos{\beta} &       0      & \sin{\beta} \\
      0      &       1      &      0      \\
-\sin{\beta} &       0      & \cos{\beta} \\
\end{pmatrix},\nonumber
\end{equation}
\begin{equation}
{\rm R_z}\equiv
\begin{pmatrix}
\cos{\alpha} & -\sin{\alpha} & 0 \\
\sin{\alpha} &  \cos{\alpha} & 0 \\
     0       &       0        & 1  
\end{pmatrix},\hspace{0.2cm}
{\rm R_\delta}\equiv
\begin{pmatrix}
\cos{\delta} &       0      & \sin{\delta} \\
      0      &       1      &      0      \\
-\sin{\delta} &       0      & \cos{\delta} \\
\end{pmatrix},\nonumber
\end{equation}
give the rotations on the $x$, $y$, and $z$ axes, respectively; $\alpha$, $\beta$, and $\gamma$ are the Euler's angles, and $\delta$ is the inclination of the loop with respect to the radial direction.
The six free parameters are the semi-minor and semi-major axes of the elliptical loop ($a_{\rm o}$ and $b_{\rm o}$), the spherical coordinates ($\theta_{\rm o}$ and $\phi_{\rm o}$) defining the center of the ellipse, $\delta$, and one of the three Euler angles ($\gamma$).
An important assumption is that the location of the radio source emitting at the minimum frequency (298.7 MHz) corresponds to the top of the loop. 
To satisfy the above condition, one of the two Euler angles was constrained by the requirement that $\beta = \theta_{\rm o}$.
As for $\alpha$, this Euler's angle was simply given by $\phi_{\rm o}$.
The six best-fit parameters were obtained by simultaneously minimizing the model parameters over their entire range of variability with respect to the positions of the radio sources. 
We adjusted these parameters until the best-fit loop model matched the morphology of the observed loop (as imaged by NRH) in projection on the plane of the sky.
Further requirements were that the locations of the radio sources from which the other two frequencies (327.0 and 360.8 MHz) were emitted along the loop were symmetrical with respect to the vertical semi-axis of the semi-ellipse.
This allowed us to better constrain the shape of the loop.

\begin{figure*}
    \centering
    \includegraphics[width=15cm]{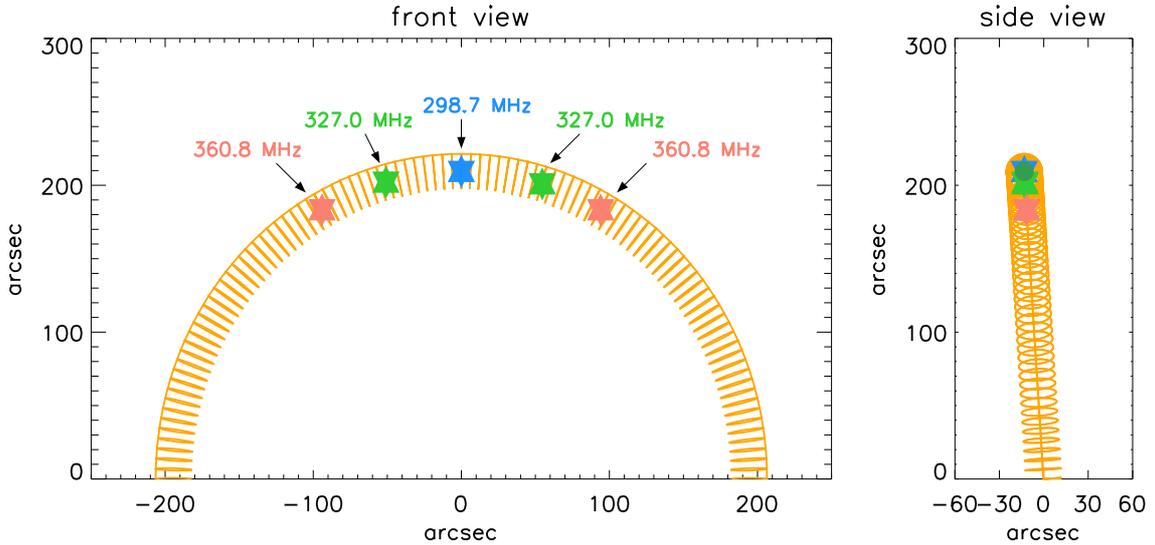}
    \caption{Schematic view of the coronal loop model together with the radio source positions as detected by the NRH. The left panel shows the front view, corresponding to the perpendicular direction to the plane of the loop, while the right panel shows the side view, corresponding to the parallel direction to the plane of the loop. }
    \label{Fig5}
\end{figure*}

The best-fit parameters we obtained with the above model are $a_{\rm o} = 0.218$ \rsun, $b_{\rm o} = 0.202$ \rsun, $\delta = 3.6$\dg, $\gamma = 30.7$\dg, ${\theta}_{\rm o} = -21.4$\dg, and ${\phi}_{\rm o} = -205.0$\dg.
The values of ${\theta}_{\rm o}$ and ${\phi}_{\rm o}$ refer, respectively, to the latitude of the center of the ellipse with respect to the solar equator and to its longitude with respect to the central meridian as seen from Earth at the time of the event.
The location on the solar surface of the reconstructed loop through which the \tu\ radio burst propagated is shown in Figure~\ref{Fig2} superposed on a simultaneous SDO/AIA image taken at 171 \AA\ and in the left panel of Figure~\ref{Fig4} as seen from the perspective of Earth.
Although the fitting procedure yielded a clear minimum in the least-square procedure, an ambiguity still remains along the line of sight such that the minimum can be obtained by also imposing $\gamma \rightarrow -\gamma$ and ${\theta}_{\rm o} \rightarrow -{\theta}_{\rm o}$.
On the other hand, it is reasonable to assume, as also expected for \tiii\ radio burst emission, that the high degree of directivity of the subrelativistic electron beams inside the loop favors emission toward the observer. 
Under this plausible assumption, it is possible to completely disentangle the above ambiguity and conclusively derive the actual topology of the coronal loop above the disk.
The right panel of Figure~\ref{Fig4} shows the location of the loop above the solar disk under a different perspective, that is, with the Sun rotated by 90\dg\ to the west.
Figure~\ref{Fig5} displays a schematic view of the loop model as seen from the front view (left panel), corresponding to the perpendicular direction to the plane of the loop, and from the side view (right panel), corresponding to the parallel direction to the plane of the loop.

\subsection{Physical properties of the coronal loop }

As already discussed in Sect. 1, \tu\ radio bursts represent a subclass of the most common \tiii\ radio bursts in that they are produced via coherent emissions induced by energetic electron beams propagating along closed magnetic coronal loops from one foot point to the other.
However, \tu\ radio bursts are not observed as frequently as \tiii\ radio bursts, even though most of the coronal magnetic flux is contained in closed-loop configurations. 
It is therefore important to identify the peculiar mechanism underlying their unexpected limited production.

Assuming electron beam acceleration with a power-law energy spectrum (e.g., \citealt{Holman2011}), the electron beams confined in the coronal loop are known to undergo an initial period of propagation before their distribution function becomes unstable (as an effect of a positive gradient in velocity space) to Langmuir waves so that it produces coherent radio emission (\citealt{Ginzburg1958}) at the fundamental or second harmonic of the plasma frequency $f_{\rm pe}$, which is related to the electron density $n_{\rm e}$ by the simple relation $f_{\rm pe} [{\rm MHz}] \approx 0.009 \sqrt{n_{\rm e} [{\rm cm}^{-3}]}$.
The electron density $n_{\rm e}$ inside the loop can therefore easily be obtained, apart from a generally unknown factor of two related to the uncertainty in identifying the correct harmonic from the observed radio dynamic spectrum.
In the following discussion, we provide evidence that the observed radiation is indeed due to second harmonic emission, meaning that $f = 2 f_{\rm pe}$, thus providing a sound estimate for the electron density $n_{\rm e}$ near the top of the loop.

Electromagnetic waves are absorbed in a plasma, because electrons oscillating in the wave fields collide and lose a part of the energy of the waves.
Collisional damping reduces the intensity of transverse waves particularly if radiated near the plasma frequency $f_{\rm pe}$.
The effect of the so-called free-free absorption of radiation between source and observer can be estimated for radiation of frequency $f$ emitted in an exponential atmosphere of scale height $H$.
The optical depths of the fundamental ($\tau_{\rm F}$) and  harmonic ($\tau_{\rm H}$) components for free-free absorption (\citealt{Robinson2000}) are given, respectively, by
\begin{equation}
\tau_{\rm F} = \left({f \over 76.3 \hspace{0.1cm} {\rm MHz}}\right)^2\left({T_{\rm e} \over 2 \hspace{0.1cm} {\rm MK}}\right)^{-3/2}{H \over 10^{10} \hspace{0.1cm} {\rm cm}},
\end{equation}
\begin{equation}
\tau_{\rm H} = \left({f \over 497 \hspace{0.1cm} {\rm MHz}}\right)^2\left({T_{\rm e} \over 2 \hspace{0.1cm} {\rm MK}}\right)^{-3/2}{H \over 10^{10} \hspace{0.1cm} {\rm cm}}.
\end{equation}
This implies that the collisional absorption depends on the observed emission frequency $f$ relative to the plasma frequency $f_{\rm pe}$ and that the observable fundamental emission is strongly suppressed above a few hundred MHz, while harmonic emission is suppressed above roughly 1 GHz.
By using typical coronal temperatures of $T_{\rm e} \approx 1-2 \cdot 10^6$ K and a typical scale height $H = 0.2$ \rsun, as obtained for example by adopting the coronal density model of \cite{Newkirk1961}, we find $\tau_{\rm F} \approx 20-60$ and $\tau_{\rm H} \approx 0.5-1.5$ near the top of the loop.
Harmonic emission is therefore much less affected than fundamental radiation, because little radiation emerges from $\tau > 3$ (e.g., \citealt{Benz2002}).
As a consequence, the flux of the fundamental emission is significantly lower than that of harmonic emission and is expected to be below the lower thresholds of typical radio instruments, irrespective of whether damping by scattering off density fluctuations is important or not.
Finally, the polarization degree of the second harmonic plasma emission is generally low, as also found for this event.
We therefore assume throughout the paper that the observed emission mechanism is the second harmonic plasma emission, 2$f_{\rm pe}$, meaning that, for example, the density at the top of the loop (where $f = f_{\rm H} = 2f_{\rm pe} \approx 300$ MHz) is estimated at about $n_{\rm e} = (f_{\rm pe} {\rm [MHz]}/0.009)^2 \approx (150/0.009)^2 \approx  3.0\cdot10^8$ cm$^{-3}$. 

For the coronal plasma in the loop to be magnetically confined, the following requirement for the plasma $\beta$ parameter, representing the ratio of thermal pressure $P = n_{\rm e}k_{\rm B}T$ to magnetic pressure $P_B$, 
\begin{equation}
\beta = {P \over P_{\rm B}} = {n_{\rm e} k_{\rm B} T \over B^2/8\pi} \ll 1,
\end{equation}
must be satisfied. 
In static loop models, it is assumed that the  volumetric heating rate is balanced by the conductive and radiative loss rate.
Assuming hydrostatic equilibrium, it is possible to apply the \cite{Rosner1978} scaling law to estimate the temperature maximum at the apex of a quasi-stationary loop structure.
This scaling law provides a unique relation between the loop maximum temperature $T$, the pressure $P$, and the size $L$, given by 
\begin{equation}
T_{\rm max} = 6.154 \cdot 10^{-4} \sqrt{n_{\rm e} L} \hspace{0.2cm}[{\rm K}], 
\end{equation}
in cgs units. 
According to the above relation, as $L \approx r_{\rm o} \pi/2 = 0.33$ \rsun\  ($r_{\rm o} = 0.21$ \rsun\ is the radius of the loop) and $n_e \approx 3.0 \cdot 10^8$ cm$^{-3}$ at the top of the loop, we obtain $T_{\rm max} \approx 1.6 \cdot 10^6$ K.
According to the classification of coronal loops given in the review of \cite{Reale2014}, the physical parameters inferred for this loop  are typical of warm giant arches.

It is known that it is very difficult to measure magnetic fields in the corona, and this can be done only in very special conditions, such as in very strong local fields (\citealt{White1991}), flare-induced coronal loop oscillations (\citealt{Nakariakov2001}), type II radio emission from shocks driven by coronal mass ejections (e.g., \citealt{Mancuso2013b,Mancuso2019}), or using Faraday rotation techniques (e.g., \citealt{Mancuso2000, Mancuso2013a}).
 Using eq. (4), we find that in order to confine the plasma contained in the loop, the magnetic field strength $B$ must largely exceed at least $\approx 1.3$ G near the top of the loop.

\subsection{Electron beam characteristics}

If a distribution function has more particles at higher velocities than at lower velocities in some region of phase space, this distribution will be unstable to waves that are in resonance with the particles. 
Due to this effect, known as bump-on-tail instability, a beam of electrons with velocities $v_{\rm b}$ much higher than the thermal speed $v_{\rm th}$ of electrons in the plasma causes the exponential growth in amplitude of Langmuir waves, which can then undergo wave--wave processes to nonlinearly excite and emit electromagnetic radiation near the local plasma frequency and its second harmonic (e.g., \citealt{Ginzburg1958,Melrose1980,Cairns1987,Robinson1993}).
The energy for the Langmuir waves to grow comes from the accelerated electron beams. 
Quasi-linear theory predicts that the system evolves to a state in which two-thirds of the initial beam kinetic energy goes to the waves, while one-third of it is retained by the electrons, whose distribution function is flattened into a plateau in velocity space (\citealt{Melrose1985}) so that the growth rate for Langmuir waves is deemed to vanish. 
This energy extraction causes a lower kinetic energy with consequent deceleration, until the beam velocity $v_{\rm b}$ reaches a threshold of about $4v_{\rm th}$ for Langmuir waves (e.g., \citealt{Benz2002,Aschwanden2005}).
Moreover, Landau damping inhibits the growth of Langmuir waves with velocities below about four times the ambient electron thermal velocity, and so we expect $v_{\rm b} \gtrsim 4v_{\rm th}$.
Another important aspect is the level of density fluctuations:  when the average level of density fluctuations becomes of the order of or larger than about $3 (v_{\rm th}/v_{\rm b})^2 $, the dynamics of the system is strongly influenced by the scattering of the waves on the density inhomogeneities (see discussion in \citealt{Krafft2015}).
In particular, compared to the homogeneous plasma case, the rate of growth of the Langmuir wave energy emitted during the bump-on-tail instability is decreased.
Therefore, the influence of density fluctuations inside the loop on the nonlinear dynamics of waves during Langmuir turbulence is also important and should, in principle,
be considered as well.

\begin{figure}[t]
    \centering
    \includegraphics[width=8.5cm]{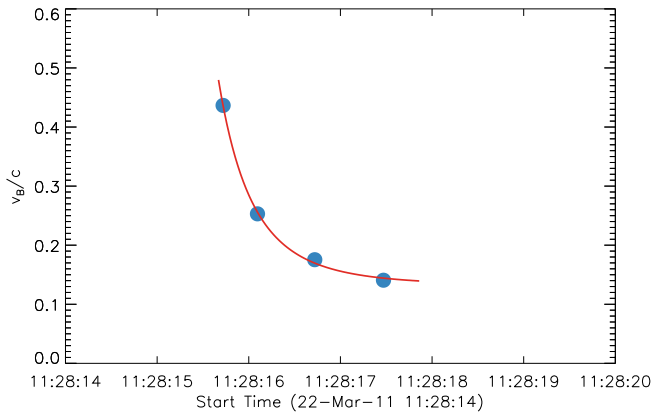}
    \caption{Temporal dependence of the electron beam speed inside the loop. The solid curve is the best-fit exponential decay to a constant level.
    }
    \label{Fig6}
\end{figure}

If we assume semi-relativistic electron beam acceleration injected with a power-law energy spectrum, we find that the beam is subject to an initial time interval of propagation in which it is not unstable to Langmuir wave production. 
On the other hand, above a given {instability distance}, the electron beams develop the bump-on-tail velocity distribution required for the generation of Langmuir waves.
According to quasi-linear simulations (\citealt{Reid2011}), this distance depends on the electron beam velocity spectral index $\gamma$, the size $d$ of the acceleration site, and the temporal injection profile. 
For the \tu\ radio burst observed in this study, its starting height, and therefore the instability distance, is at least 0.2 \rsun.
This distance depends largely on the spectral index of the injected electron beam and the longitudinal extent of the acceleration region (\citealt{Reid2011,Reid2014}).
Assuming a typical acceleration region size of $d = 0.015$ \rsun\ and an acceleration region height of $h_{\rm acc} = 0.07$  \rsun\ (e.g., \citealt{Reid2017}), in case of instantaneous injection we expect the radio emission to start at $h = d\gamma + h_{\rm acc} \approx 0.2$ \rsun, thus yielding $\gamma \approx 9$.
On the other hand, the latter can only be considered as an upper limit. In fact, the instability distance is also governed by the temporal injection profile that plays a significant role if $v_{\rm b}\tau_{\rm inj}>d$, where $\tau_{\rm inj}$ is the injection time, in which case $h = (d+v_{\rm b}\tau_{\rm inj})\gamma + h_{\rm acc}$.

A rough lower limit to the electron beam velocity $v_{\rm b}$ can be estimated by considering  the deflection time $\tau_{\rm D}$ by Coulomb collisions with thermal particles (e.g., \citealt{Hughes1963,Elgaroy1972}), which is given by
\begin{equation}
    \tau_{\rm D} = 3.1 \times 10^{-20} {v_{\rm b}^3 \over n_{\rm e}} \hspace{0.2cm}[{\rm s}],
\end{equation}
in cgs units.
Given the total duration of the \tu\ radio burst ($\approx 2.5$ s), and assuming $n_{\rm e} \approx 3\cdot 10^8$ cm$^{-3}$ from its turnover frequency (see Figure~\ref{Fig5}), the minimum velocity of the electron beam obtained with the above formula is $v_{\rm b} \approx 0.1c$, where $c$ is the speed of light. 
If we take into account the correct geometry of the loop as determined from our model and the electron beam velocity as derived from the radio observations, by fitting the result with an exponential decay to a constant level, 
we obtain the speed of the exciter as
\begin{equation}
v_{\rm b}(t) = \left[0.14+21.87\cdot{\rm exp}(-t/0.40)\right]c, 
\end{equation}
which appears to be, as expected, above the estimated lower limit.
Figure~\ref{Fig6} shows how Coulomb collisions with thermal electrons rapidly affect, with an e-folding time of 0.40 s, the propagation of the electron beam inside the loop.

Considering the maximum loop temperature $T_{\rm max}$ determined in the previous analysis ($T_{\rm max} \approx 1.6 \cdot 10^6$ K), corresponding to an electron thermal velocity $v_{\rm th} = \sqrt{2 k_{\rm B} T/m_{\rm e}} \approx 0.02 c$, this would imply a ratio $v_{\rm b}/v_{\rm th}$ decreasing from about 19 to 6, that is, just above the threshold velocity needed for the generation of Langmuir waves. 
The corresponding electron energies therefore decrease from about 60 keV to 5 keV after 2 s.
Assuming $d = 0.015$ \rsun, for $v_{\rm b} \approx 0.45c$ the injection time can be as low as $\tau_{\rm inj} < d/v_{\rm b} \approx 0.07$ s, implying an even lower spectral index ($\gamma \approx 5$) than the one previously determined.
As a consequence, our analysis clearly supports the evidence that the \tu\ radio burst extinguished because the electron beams related to this coherent radio emission traveled through a denser plasma in the coronal loop, thus tending to be isotropized faster by Coulomb collisions, until they reached a threshold velocity that did not meet the necessary criterion for beam-driven plasma emission (that is $v_{\rm B} \gtrsim 4v_{\rm th}$) due to bump-on-tail instability.

\section{Discussion and conclusions}

It has been suggested that for radio emission to be generated on closed magnetic fields, the loop needs to be long enough for a power-law-accelerated electron beam to become unstable to Langmuir waves (see \citealt{Reid2017}). 
It follows that the \tu\ radio emission is supposed to occur far away from the beam injection region, so that the accelerated electrons actually trace the upper part of coronal loops.
Moreover, the higher density and lower density gradient of the plasma inside the loop concur in reducing the  growth rate of the Langmuir wave.
As a consequence, the beam needs to be dense enough for the timescale of Langmuir wave growth and its successive conversion to radio waves to be shorter than the electron propagation timescale.
The above conditions therefore result in a stringent set of preconditions for both the propagating electron beams and the background loop plasma parameters in order to produce \tu\ radio emission over the more common \tiii\ radio emission.
This also implies that small loops should not have sufficient heights for the required instability to occur.

In this study, we present the analysis of a \tu\ radio burst observed on 2011 March 22 by the CALLISTO ground-based radio spectrograph.
Radio imaging data from the NRH at three different frequencies (298.7, 327.0, and 360.8 MHz) were used in conjunction with a novel modeling approach to infer the 3D topology of the loop inside which the subrelativistic electron beams
propagated, producing the coherent radio emission.
The confined coronal structure is identified as an almost vertical and semi-circular inclined coronal loop with a height of about 0.21 \rsun\ located  behind the eastern solar limb as seen from Earth.
The velocity of the exciting electron beam is found to be decreasing with a power-law trend from $\approx c/2$ to $c/7$, consistent with the interpretation of electron beams being instantaneously injected with a spectral index $\gamma \approx 9$, decelerated, and losing their energy with an e-folding time of 0.40 s, and therefore being isotropized along the way due to Coulomb collisions in the loop. 
Assuming hydrostatic equilibrium, we were also able to constrain the maximum loop temperature ($T_{\rm max} \approx 1.6 \cdot 10^6$ K), while in order to confine the plasma contained in the loop, the magnetic field strength $B$ was constrained to largely exceed at least $\approx 1.3$ G near the top of the loop.

Our analysis supports the view that the \tu\ radio burst extinguished because the injected electron beams related to this coherent radio emission traveled through a denser plasma in the coronal loop, thus tending to be isotropized faster by Coulomb collisions, until they reached a threshold velocity that did not meet the necessary criterion for beam-driven plasma emission (that is, $v_{\rm b} \gtrsim 4v_{\rm th}$) due to bump-on-tail instability.
The size of the loop as inferred in this work should imply, in light of the findings of previous studies, that its dimensions could have been just sufficient to excite the observed \tu\ radio burst and that some other major unknown factor may actually concur in the ignition of coherent radiation from a small loop.

Finally, we would like to point out that although several coronal loops of various shapes, heights, and orientations are visible in the vicinity of the coronal region under study (see Figure~\ref{Fig2}), we were not able to reliably identify an exact counterpart of the modeled loop as outlined by the NRH radio sources in any of the available SDO/AIA EUV channels.
This is problematic, but could be attributed to a selection effect due to the fact that the low  density of the confined plasma in this loop could be difficult to detect (see \citealt{Reale2014}) and that the presence of denser foreground coronal structures might have hindered its detection. 
On the other hand, as the modeled loop was positioned behind the east limb, as indicated from Figure~\ref{Fig4}, the presence of strong refraction and scattering due to tangential propagation of the radio waves through the intervening dense inhomogeneous structures along the line of sight to the observer is another concrete possibility.
In fact, it is known that in general, due to refraction in the corona, the apparent location of radio sources near the limb tends to be shifted toward the solar disk center, while scattering tends to shift their positions farther than the disk center. Anisotropic scattering from nonradial structures can further contribute in an effective manner to nonradial shifts in the observed positions of the radio sources. 
Indeed, \cite{sharma2020} found clear evidence for the presence of significant propagation effects, including anisotropic scattering, in their radio observations with the Murchison Widefield Array up to 240 MHz.
Moreover, wave ducting can cause shifts in the radio source position under specific conditions of the emission angle with the coronal loop boundaries (\citealt{Duncan1979}). 
Notwithstanding the above caveats, at the relatively high NRH radio frequencies considered in this work, we still expect the adverse contribution from scattering or refraction to play, in general, a less important role than observed. 
However, taking into account the much lower spatial resolution of NRH with respect to the SDO/AIA instrument and the recent findings of \cite{sharma2020}, it is still possible that radio wave propagation effects have somehow hindered the correct identification of the loop in the EUV images.

In summary, we show that the analysis of \tu\ radio bursts is a powerful remote sensing diagnostic tool for both the electron beams and the confined arched
plasma structures they travel through. 
We demonstrate that the use of high-resolution radio heliograph images of \tu\ radio bursts in conjunction with their radio dynamic spectra  can be used to reconstruct the 3D morphology of the coronal loop inside which they propagate, alleviating the need for triangulation techniques. 
Moreover, the analysis of such events is a powerful and unique remote sensing diagnostic tool for constraining both the dynamics of the exciting electron beams and the physical parameters of the coronal loops in which they propagate.

\begin{acknowledgements} 
We thank the referee for carefully reading our paper and providing useful comments and suggestions. 
We also thank the NRH and SDO teams, and the e-CALLISTO network for providing open data access. The NRH is supported by the Observatoire de Paris, CNRS and Région Centre. 
F. F. is supported through the Metis programme funded by the Italian Space Agency (ASI) under the contracts to the co-financing National Institute of Astrophysics (INAF): Accordo ASI-INAF n. 2018-30-HH.0.

\end{acknowledgements}

\bibliographystyle{aa}
\bibliography{biblio}

\end{document}